\documentclass[12pt]{article}
\usepackage{tikz}
\usepackage{amsmath,amssymb}
\usepackage{bbm}  
\usepackage{fancyhdr}
\usepackage[utf8]{inputenc}    
\usepackage{fullpage}
\usepackage{float}
\usepackage{authblk}
\usepackage{color}
\usepackage{caption}
\usepackage{tikz, tikz-qtree}

\let\OLDthebibliography\thebibliography
\renewcommand\thebibliography[1]{
  \OLDthebibliography{#1}
  \setlength{\parskip}{0pt}
  \setlength{\itemsep}{0pt plus 0.3ex}
}

\usepackage{tikz}
\usetikzlibrary{patterns}
\usetikzlibrary{calc,arrows,positioning}
\usetikzlibrary{arrows,shadows}
\usepackage{circuitikz} 
\usetikzlibrary{decorations.pathmorphing}	
\usetikzlibrary{decorations.markings}
\author[1,2]{Etienne Granet}
\author[1,2,3]{Jesper Lykke Jacobsen}
\author[1,4]{Hubert Saleur}
\affil[1]{Institut de Physique Th\'eorique, Paris Saclay, CEA, CNRS, 91191 Gif-sur-Yvette, France}
\affil[2]{Laboratoire de Physique de l'Ecole Normale Sup\'erieure, ENS, Universit\'e PSL, CNRS, Sorbonne Universit\'e, Universit\'e Paris-Diderot, Sorbonne Paris Cit\'e, Paris, France}
\affil[3]{Sorbonne Universit\'e, \'Ecole Normale Sup\'erieure, CNRS, Laboratoire de Physique (LPENS), 75005 Paris, France}
\affil[4]{USC Physics Department, Los Angeles CA 90089, USA}

\title{Analytical results on the Heisenberg spin chain in a magnetic field}
\date{}

\begin{document}
\maketitle
\begin{abstract}
We obtain the ground state magnetization of the Heisenberg and XXZ spin chains in a magnetic field $h$ as a series in $\sqrt{h_c-h}$, where $h_c$ is the smallest field for which the ground state is fully polarized. All the coefficients of the series can be computed in closed form through a recurrence formula that involves only algebraic manipulations. For some values of the anisotropy parameter $\Delta$ the expansion is numerically observed to be convergent in the full range $0\leq h\leq h_c$.

To that end we express the free energy at mean magnetization per site $-1/2\leq \langle \sigma^z_i\rangle\leq 1/2$ as a series in $1/2-\langle \sigma^z_i\rangle$ whose coefficients can be similarly recursively computed in closed form. This series converges for all $0\leq \langle \sigma^z_i\rangle\leq 1/2$. The recurrence is nothing but the Bethe equations when their roots are written as a double series in their corresponding Bethe number and in $1/2-\langle \sigma^z_i\rangle$. It can also be used to derive the corrections in finite size, that correspond to the spectrum of a free compactified boson whose Luttinger parameter can be expanded as a similar series.

The method presumably applies to a large class of models: it also successfully applies to a case where the Bethe roots lie on a curve in the complex plane.
\end{abstract}

\section{Introduction}
The Heisenberg spin chain and its anisotropic XXZ version are paradigmatic models of quantum interacting systems that exhibit long-range correlations. Their Hamiltonian with periodic boundary conditions,  anisotropy $\Delta$ and magnetic field $h$ is
\begin{equation}
H=\sum_{i=1}^N \sigma^x_i\sigma^x_{i+1}+ \sigma^y_i\sigma^y_{i+1}+ \Delta (\sigma^z_i\sigma^z_{i+1}-\tfrac{1}{4})-h\sigma^z_i\,.
\end{equation}
 The reason for such an interest is their integrability: the Bethe ansatz \cite{bethe,orbach,yangyang1,sklyanin} enables, in principle, to compute energy levels or form factors or spin-spin correlations exactly (see e.g. \cite{slavnov} for a recent review). However, the simple and physically relevant question of the magnetization acquired by the spin chain when put in an external magnetic field is still unsolved \cite{griffiths,yangyang3}. It is equivalent to finding the free energy of the chain at any mean magnetization per site $-1/2\leq \langle \sigma^z_i\rangle\leq 1/2$. Although a root density approach allows to find a closed form expression for the free energy at zero magnetization \cite{hulthen,yangyang}, it does not permit to derive a magnetization-dependent expression. In this approach indeed, one needs to solve a so-called Wiener-Hopf equation on a finite interval \cite{griffiths,korepin,essler,gaudin,cabrahoneckerpujol}
 \begin{equation}
 \label{eq:whopf}
\rho(x)+\int_{-\Lambda}^\Lambda \rho(t)r'(x-t)dt=s'(x)\,,
\end{equation}
where the unknown is $\rho$, while $r$ and $s$ are given, and $0<\Lambda<+\infty$. The equation holds for all values of $x$ and no periodicity of $s$ and $r$ is assumed (contrarily to the case $\Delta>1$ where a similar equation holds at zero magnetic field, but with periodic $s$ and $r$). This equation is equivalent to a two-dimensional classical Wiener-Hopf equation, for which no generic solution is known \cite{noble, danielezich}. Only a first-order perturbation around zero magnetic field could be performed in this approach for various physical quantities \cite{yangyang,bogoliubovizerginkorepin,essler,japaridzenersesyan}. Since the first appearance of this approach in \cite{griffiths}, apart from some works at finite temperature \cite{takahashi2}, no alternative method seems to have been tried.\\

In this paper we present a different approach that expresses the free energy as a power series in $1/2- \langle \sigma^z_i\rangle$, whose coefficients can be recursively computed in an explicit closed form with only algebraic manipulations. It leads to an expansion of the acquired magnetization of the chain in a magnetic field $h$, in terms of $\sqrt{h_c-h}$ where $h_c$ is the smallest field for which the ground state is fully polarized. The calculation does not use root densities and is actually quite simple. The Bethe roots are written as a double power series in their corresponding Bethe number and in $1/2- \langle \sigma^z_i\rangle$. The Bethe equations are then nothing but a recurrence relation on these coefficients. Surprisingly, this approach has not been tried before. Only the first coefficient in front of $\sqrt{h_c-h}$ was known \cite{bonnerfisher,hodgsonparkinson,korepin}, or the first two ones in similar models for other physical quantities \cite{konikfendley}.\\

The method presumably works for a large class of models. As an example we present another case where the Bethe roots lie on a curve in the complex plane, and derive similarly a power series for the free energy.

\section{The Bethe equations as a recurrence relation}
We consider Bethe equations written in logarithmic form for $k$ roots $\lambda_i$ ($k$ even):
\begin{equation}
s(\lambda_i)=\frac{I_i}{N}+\frac{1}{N}\sum_{j=1}^kr(\lambda_i-\lambda_j)\,,
\end{equation}
with $I_i=-k/2+1/2,...,k/2-1/2$ increasing half-integers. $s$ and $r$ are functions depending on the model, that are assumed to be odd and satisfy $s'(0)\neq 0$. Our goal is to compute the free energy
\begin{equation}
\label{eq:deffe}
F=-\frac{2\pi}{N}\sum_{i=1}^k s'(\lambda_i)\,,
\end{equation}
in the limit $N\to\infty$, with $k/N$ finite\footnote{The fact that the free energy is defined in terms of the same function $s$ as in the Bethe equations is clearly not a constraint of the method, and it could be defined in terms of another function $f$ as well.}. We set
\begin{equation}
\alpha_i=\frac{I_i}{N}\,,\qquad m=\frac{k}{N}=\frac{1}{2}- \langle \sigma^z_i\rangle\,,
\end{equation}
and first assume that all the $\lambda_i$ and $\lambda_i-\lambda_j$ lie within the radius of convergence of $s$ and $r$ respectively as a power series around $0$. The starting point is to write each $\lambda_i$ as a power series in $\alpha_i$ and $m$:
\begin{equation}
\lambda_i=\sum_{a,b\geq 0} \alpha_i^a m^b c_{ab}\,,
\end{equation}
with $c_{ab}$ some coefficients to determine. For convenience, we will use the notations for $n\geq 1$
\begin{equation}
\lambda_i^n=\sum_{a,b\geq 0} \alpha_i^a m^b c_{ab}^{[n]}\,,\qquad \text{where }  c_{ab}^{[n]}=\sum_{\substack{a_1+...+a_{n}=a\\ b_1+...+b_{n}=b}} c_{a_1b_1}...c_{a_{n}b_n}\,.
\end{equation}
By definition we set $c_{00}^{[0]}=1$, the other coefficients $c_{ab}^{[0]}$ being zero. Expanding $s(\lambda_i)$ yields
\begin{equation}
s(\lambda_i)=\sum_{a,b\geq 0}\alpha_i^a m^b\sum_{n\geq 0}\frac{ s^{(n)}(0)}{n!}c_{ab}^{[n]}\,.
\end{equation}
Similarly for $r(\lambda_i-\lambda_j)$:
\begin{equation}
r(\lambda_i-\lambda_j)=\sum_{n\geq 0} \frac{r^{(n)}(0)}{n!}\sum_{q=0}^{n}(-1)^q{n \choose q} \sum_{a_1,a_2,b_1,b_2\geq 0}\alpha_j^{a_1} m^{b_1}\alpha_i^{a_2} m^{b_2} c_{a_1b_1}^{[q]}c_{a_2b_2}^{[n-q]}\,.
\end{equation}
Now one has for even $a$
\begin{equation}
\label{eq:sum}
\frac{1}{N}\sum_{j=1}^k \alpha_j^a=\frac{2}{N^{a+1}}\sum_{j=1}^{k/2}(j-\tfrac{1}{2})^a=\frac{m^{a+1}}{2^{a}(a+1)}+O(N^{-1})\,,
\end{equation}
while for $a$ odd it is zero. This gives
\begin{equation}
\frac{1}{N}\sum_{j=1}^k r(\lambda_i-\lambda_j)=\sum_{a,b\geq 0}\alpha_i^a m^b\sum_{\substack{a_1+b_1+b_2+1=b\\ a \text{ even}}}\sum_{n\geq 0}\frac{r^{(n)}(0)}{n!}\sum_{q=0}^{n} (-1)^q {n \choose q}\frac{c_{a_1b_1}^{[q]} c_{a_2b_2}^{[n-q]}}{2^{a_1} (a_1+1)}\,.
\end{equation}
Since $c_{00}=0$, each $c^{[n]}_{ab}$ depends on $c_{a'b'}^{[n-1]}$ with $a'\leq a$, $b'\leq b$, and at least $a'<a$ or $b'<b$. Thus one gets the following recurrence relation
\begin{equation}
\label{eq:recursive}
c_{ab}=\frac{1}{s'(0)}\left(\sum_{\substack{a_1+b_1+b_2+1=b\\ a_1 \text{ even}}} \left(\sum_{n\geq 0}\frac{r^{(n)}(0)}{n!}\sum_{q=0}^{n}(-1)^q {n \choose q}\frac{c_{a_1b_1}^{[q]} c_{ab_2}^{[n-q]}}{2^{a_1} (a_1+1)}\right)-\sum_{n\geq 2} \frac{c^{[n]}_{ab}s^{(n)}(0)}{n!} \right)\,,
\end{equation}
with the initial conditions $c_{00}=0,c_{10}=\tfrac{1}{s'(0)}$. Since $c_{ab}^{[n]}=0$ for $n$ large enough at fixed $a,b$, all the sums on the right-hand side are in fact finite sums. Note that $c_{ab}$ depends on $c_{a'b'}$ with either (i) $a'\leq \max (b-1,a)$ and $b'\leq b-1$, and if $a'>a$ then $a'+b'\leq b-1$; or (ii) $b'=b$ and $a'<a$. In both cases one has $a'+b'<a+b$. Thus one can compute the $c_{ab}$'s by computing $c_{d-b,b}$ for increasing $b$'s, and then increasing $d$'s.

The free energy $F(m)$ is now
\begin{equation}
F(m)=-\frac{2\pi}{N}\sum_{i=1}^k\sum_{n\geq 0}\frac{s^{(n+1)}(0)}{n!}\sum_{a,b\geq 0}\alpha_i^a m^b c_{ab}^{[n]}\,.
\end{equation}
This gives the expansion
\begin{equation}
\label{eq:freeenergy}
\begin{aligned}
&F(m)=\sum_{a\geq 0}m^a f_a\\
&\text{with }\quad f_a=-\frac{\pi}{2^{a-1}}\sum_{\substack{b=0\\a-b \text{ odd}}}^{a} \frac{2^{b+1}}{a-b} \sum_{n\geq 0}\frac{s^{(n+1)}(0) }{n!}c^{[n]}_{a-b-1,b}\,.
\end{aligned}
\end{equation}

The calculation assumes that $\lambda_i$ and $\lambda_i-\lambda_j$ lie within the radius of convergence of $s$ and $r$; however it may happen that the resulting series $F(m)$ has a larger radius of convergence. If the free energy has the property of being expandable in a power series in all the physical range of $m$, then the two expressions must coincide for all $m$.


\section{Examples}

\subsection{The Heisenberg spin chain}
\subsubsection*{The free energy}
For the Heisenberg spin chain one has the functions
\begin{equation}
s(\lambda)=\frac{1}{\pi}\arctan 2\lambda\,,\qquad r(\lambda)=\frac{1}{\pi}\arctan\lambda\,,
\end{equation}
with
\begin{equation}
s^{(2n+1)}(0)=\frac{(-1)^n (2n)! 2^{2n+1}}{\pi}\,,\qquad r^{(2n+1)}(0)=\frac{(-1)^n (2n)!}{\pi}\,.
\end{equation}

Applying \eqref{eq:recursive} and plugging the $c$'s into \eqref{eq:freeenergy}, one directly finds the expansion
\begin{equation}
\label{eq:fexxx}
\begin{aligned}
&F(m)=-4m+\frac{\pi^2}{3}m^3+\frac{\pi^2}{3}m^4+\left(-\frac{\pi^4}{60}+\frac{\pi^2}{4} \right) m^5+\left(-\frac{\pi^4}{36}+\frac{\pi^2}{6} \right)m^6\\
&+\left( \frac{\pi^6}{2520}-\frac{\pi^4}{36}+\frac{5\pi^2}{48}\right)m^7+\left(\frac{\pi^6}{1440}-\frac{\pi^4}{48}+\frac{\pi^2}{16} \right)m^8\\
&+\left(-\frac{\pi^8}{181440}+\frac{23\pi^6}{34560}-\frac{7\pi^4}{576}+\frac{7\pi^2}{192} \right) m^9+O(m^{10})\,,
\end{aligned}
\end{equation}
and it can be readily continued from the recurrence relation. For example we computed the first $40$ terms analytically with a computer (though we only present here the first few terms), and the first $100$ terms numerically. We numerically observe that the radius of convergence of this series is $1/2$, i.e. it converges for all physical values of $m$. The expansion converges exponentially fast with the number of terms when $m<1/2$, and quadratically (possibly with a multiplicative logarithmic correction) for $m=1/2$. See Table \ref{tab:xxx} for a comparison with the direct numerical solution of the Bethe roots in finite size $N$ (the extrapolation is linear in $1/N$ using sizes $3000$ and $5000$).

\begin{center}
\begin{tabular}{|c|c|c|c|c|c|}
\hline
 & $m=0.05$ & $m=0.1$ &$m=0.2$ &$m=0.4$&$m=0.5$\\
 \hline
$N=1000$ & $0.19956813$ & $0.39637426$& $0.76823447$&$1.30357831$&$1.38629600$\\
 \hline
$N=3000$ & $0.19956797$ & $0.39637394$ & $0.76823379$&$1.30357701$&$1.38629454$\\
 \hline
 $N=5000$& $0.19956796$ & $0.39637391$ & $0.76823374$&$1.30357690$&$1.38629442$\\
 \hline
Extrapolation & $0.19956794$ & $0.39637388$ & $0.76823366$&$1.30357675$&$1.38629425$\\
 \hline
$20$-term expansion & $0.19956795$ & $0.39637390$&$0.76823371$ &$1.30357732$&$1.38641192$\\
\hline
\end{tabular}
\captionof{table}{The free energy $-F(m)$ of the Heisenberg model for different magnetizations.}\label{tab:xxx}
\end{center}

\begin{figure}[H]
 \begin{center}
\includegraphics[scale=0.7]{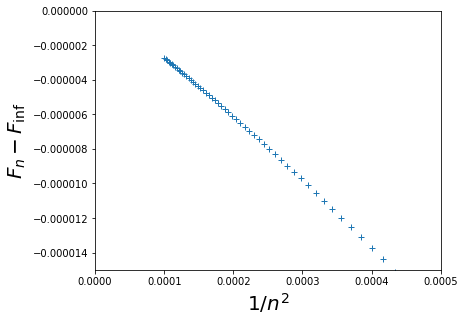} 
\end{center}
\caption{Difference between the series $F_n$ truncated after the $n$th term and its known exact limit value $F_{\rm inf}=F(1/2)$, as a function of $1/n^2$.}
 \label{figKk}
\end{figure}

For $m=1/2$, we already know from the root density approach that $F(1/2)=-2\log 2$. It is a very non-trivial check of the validity of our approach that \eqref{eq:fexxx} converges to this value, see Figure \ref{figKk}. It is worth noting that although \eqref{eq:recursive} assumes that the Bethe roots and their differences lie within the radius of convergence of $s$ and $r$ (which are $1/2$ and $1$), the resulting series still converges at some values of $m$ for which this property is not true anymore.

\subsubsection*{The acquired magnetization in a magnetic field}
Let us denote now $-1/2\leq \sigma\leq 1/2$ the mean magnetization per site of a state, i.e. the value of $\langle \sigma_i^z\rangle$ where $\sigma^z_i$ is the spin operator in the $z$ direction at site $i$, on a spin chain with $N$ sites. As already mentioned, it is related to $m$ through
\begin{equation}
\sigma=\frac{1}{2}-m\,.
\end{equation}
The energy $e_m$ of the ground state of the sector with magnetization $\sigma$ is
\begin{equation}
e_m=F(m)-(1/2-m)h\,.
\end{equation}
If the minimum of $e_m$ is obtained for $0<m^*<1/2$ then
\begin{equation}
m^*=(-F')^{-1}(h)\,.
\end{equation}
For $h=4$ one sees that $m^*=0$ and so for $h>h_c=4$ the ground state is fully polarized $\sigma^*\equiv 1/2-m^*=1/2$. For $h<h_c$ one has the following expansion in $\sqrt{h_c-h}$ of the acquired magnetization $\sigma^*$ of the chain:
\begin{equation}
\label{eq:magxxx}
\begin{aligned}
\sigma^*=&\frac{1}{2}-\frac{1}{\pi}(h_c-h)^{1/2}+\frac{2}{3\pi^2}(h_c-h)-\left(\frac{35}{72\pi^3}+\frac{1}{24\pi} \right)(h_c-h)^{3/2}+\left(\frac{10}{27\pi^4}+\frac{1}{12\pi^2} \right)(h_c-h)^2\\
&-\left(\frac{1001}{3456\pi^5}+\frac{7}{64\pi^3}+\frac{3}{640\pi} \right)(h_c-h)^{5/2}+\left(\frac{56}{243\pi^6}+\frac{10}{81\pi^4}+\frac{1}{72\pi^2} \right)(h_c-h)^3\\
&-\left(\frac{46189}{248832\pi^7}+\frac{3575}{27648\pi^5}+\frac{123}{5120\pi^3}+\frac{5}{7168\pi} \right)(h_c-h)^{7/2}\\
&+\left(\frac{110}{729\pi^8}+\frac{7}{54\pi^6}+\frac{13}{384\pi^4}+\frac{103}{40320\pi^2} \right) (h_c-h)^4+O((h_c-h)^{9/2})\,.
\end{aligned}
\end{equation}
The expansion can be continued as far as desired (we computed it exactly until order $(h_c-h)^{20}$). Only the first term $-\frac{1}{\pi}(h_c-h)^{1/2}$ was already known \cite{bonnerfisher,hodgsonparkinson,korepin}. In the case of an integer-spin chain in a magnetic field, only the first two terms of a similar expansion for quantities such as the susceptibility were computed by different means \cite{konikfendley}.\\

The two expansions \eqref{eq:magxxx} and \eqref{eq:fexxx} have not been computed before. The standard treatment of the problem was to introduce a root density $\rho$ and to try to solve the Bethe equations rewritten with this root density, that is equation \eqref{eq:whopf} where $\Lambda$ is the largest Bethe root, that depends on the magnetization of the ground state and is such that $\int_{-\Lambda}^\Lambda\rho=m$. Only an expansion at leading order in $h$ for e.g. $\Lambda$ or the critical exponents was done in \cite{yangyang,bogoliubovizerginkorepin} (see also \cite{essler} for a more recent description in case of the Hubbard model). On the field theory side, an expansion in $h$ to the first few orders was done in \cite{lukyanov}. Our result differs from these since it is an expansion in $\sqrt{h_c-h}$ where all the coefficients can be recursively computed with an explicit recurrence formula.

\subsection{\label{sec:xxz}The XXZ spin chain}
For the $XXZ$ spin chain with anisotropy parameter $-1<\Delta<1$ one has
\begin{equation}
\label{eq:sandr}
s(\lambda)=\frac{1}{\pi}\arctan\left( \sqrt{\frac{1+\Delta}{1-\Delta}}\tanh\lambda\right)\,,\qquad r(\lambda)=\frac{1}{\pi}\arctan \left( \frac{\Delta}{\sqrt{1-\Delta^2}}\tanh\lambda\right)\,,
\end{equation}
and the free energy is defined with a multiplicative factor $\sqrt{1-\Delta^2}$ in \eqref{eq:deffe} to match usual conventions. Applying the same recipe as before one finds
\begin{equation}
\label{eq:finfgamma5}
\begin{aligned}
&F(m)=-2(1+\Delta) m +\frac{\pi^2}{3}m^3+\frac{2\pi^2\Delta}{3(1+\Delta)}m^4-\frac{4\pi^2}{(1+\Delta)^2}\left(\frac{\pi^2}{240}+\frac{\pi^2\Delta}{120}+\Delta^2 \left(\frac{\pi^2}{240}-\frac{1}{4} \right) \right)m^5\\
&-\frac{8\pi^2 \Delta}{(1+\Delta)^3}\left(\frac{\pi^2}{90}+\frac{\pi^2\Delta}{120}+\Delta^2\left(\frac{\pi^2}{120}-\frac{1}{6} \right) \right) m^6+O(m^7)\,.
\end{aligned}
\end{equation}

The application of the recurrence relation is not more costly than in the case $\Delta=1$, so a $100$-term expansion is reachable.
As in the isotropic case, the radius of convergence of this series is observed to be $1/2$. The value $F(1/2)$ can be computed exactly with root densities:
\begin{equation}
F(1/2)=-\sin\gamma \int_{-\infty}^{\infty}\frac{\sinh((\tfrac{\pi}{2}-\tfrac{\gamma}{2})x)^2}{\sinh(\tfrac{\pi x}{2})(\sinh(\tfrac{\pi x}{2})+\sinh((\tfrac{\pi}{2}-\gamma)x))}dx\,,\quad \text{with }\Delta=\cos\gamma\,.
\end{equation}

\begin{figure}[H]
 \begin{center}
\includegraphics[scale=0.7]{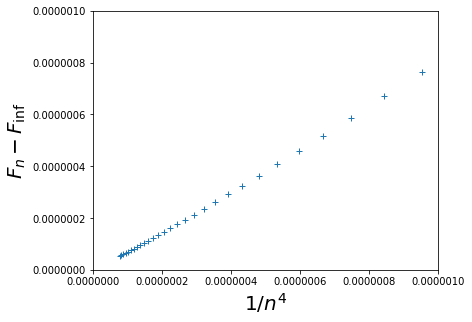} 
\end{center}
\caption{Difference between the series $F_n$ truncated after the $n$th term and its known exact limit value $F_{\rm inf}=F(1/2)$, as a function of $1/n^4$, for $\Delta=\cos \pi/5$.}
 \label{figinfgamma5}
\end{figure}

See Table \ref{tab:xxz} and Figure \ref{figinfgamma5} for the numerical verification of this expansion and of its radius of convergence, for $\Delta=\cos\pi/5$. 

\begin{center}
\begin{tabular}{|c|c|c|c|c|c|}
\hline
 & $m=0.05$ & $m=0.1$ &$m=0.2$ &$m=0.4$&$m=0.5$\\
 \hline
$N=1000$ & $0.18406350$ & $0.36172387$& $0.69569657$&$1.16699132$&$1.23606951$\\
 \hline
$N=3000$ & $0.18166924$ & $0.35938627$ & $0.69361200$&$1.16608443$&$1.23606814$\\
 \hline
 $N=5000$& $0.18119035$ & $0.35891867$ & $0.69319488$&$1.16590255$&$1.23606804$\\
 \hline
Extrapolation & $0.18047201$ & $0.35821726$ & $0.69256920$&$1.16562972$&$1.23606787$\\
 \hline
$18$-term expansion & $0.18047200$ & $0.35821721$&$0.69256908$ &$1.16562930$&$1.23605814$\\
\hline
\end{tabular}
\captionof{table}{The free energy $-F(m)$ of the XXZ spin chain for different magnetizations, for $\Delta=\cos\pi/5$.}\label{tab:xxz}
\end{center}

Here the critical magnetic field is $h_c=2(1+\Delta)$ \cite{korepin}. The expression for the acquired magnetization is
\begin{equation}
\label{eq:mag}
\begin{aligned}
\sigma^*=&\frac{1}{2}-\frac{1}{\pi}(h_c-h)^{1/2}+\frac{4\Delta}{3\pi^2(1+\Delta)} (h_c-h)\\
&-\frac{1}{72\pi^3(1+\Delta)^2}\left(3\pi^2+6\pi^2\Delta+\left(140+3\pi^2 \right)\Delta^2 \right)(h_c-h)^{3/2}\\
&+\frac{\Delta}{135\pi^4(1+\Delta)^3}\left(9\pi^2+63\pi^2\Delta+(400+18\pi^2)\Delta^2\right)(h_c-h)^2+O((h_c-h)^{5/2})\,.
\end{aligned}
\end{equation}
Only the first term $-\frac{1}{\pi}(h_c-h)^{1/2}$ was already known \cite{korepin,katsura}. 

\subsection{Radius of convergence of the series $\sigma^*(\sqrt{h_c-h})$}
\begin{figure}[H]
 \begin{center}
\includegraphics[scale=0.7]{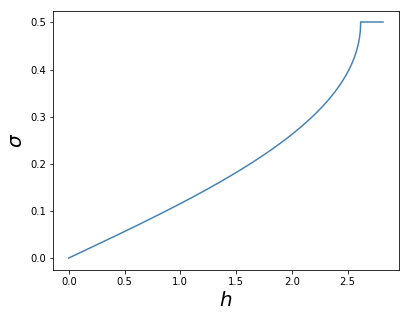} 
\end{center}
\caption{Acquired magnetization $\sigma^*$ as a function of the magnetic field $h$, using an expansion with $60$ terms, with $\Delta=\cos2\pi/5$.}
 \label{fig:mag}
\end{figure}

Although we have strong arguments in favour of the convergence of $F(m)$ as a series in $m$ in the full physical range $0\leq m\leq1/2$, this does not ensure that its Legendre transform $\sigma^*(\sqrt{h_c-h})$ has a similar property, i.e. that $\sigma^*$ as a series in $\sqrt{h_c-h}$ converges in the range $0\leq h\leq h_c$. Let us denote $h_r$ (that depends on $\Delta$) the smallest field for which this series converges in $h_r<h\leq h_c$.\\
Let us first look at the free fermion case $\Delta=0$. In this case $r=0$ in \eqref{eq:sandr}, so that the free energy $F(m)$ can be computed exactly
\begin{equation}
F(m)=-2\pi \int_{-m/2}^{m/2}s'(s^{-1}(x))dx\,.
\end{equation}
From this one deduces $h_c=2$ and
\begin{equation}
\sigma^*(\sqrt{h_c-h})=\frac{1}{2}-\frac{2}{\pi}\arctan \sqrt{\frac{h_c-h}{h_c+h}}\,,
\end{equation}
which actually has a radius of convergence $\sqrt{2h_c}$, so that the expansion converges for $-h_c<h\leq h_c$ and $h_r=-h_c$. For regularity reasons, it ensures that we must have $h_r<0$ for $\Delta$ in a neighbourhood of $0$. A strong check of $h_r<0$ is to evaluate the series at $h=0$, where one must have $\sigma^*=0$. For values of $\Delta$ not close to $1$, this is indeed observed, see Figure \ref{fig:mag} for $\Delta=\cos 2\pi/5$ where the acquired magnetization is of order $10^{-7}$ at zero magnetic field, taking into account $60$ terms in the expansion. At least for $\Delta \lessapprox 0.6$ this convergence at $h=0$ is observed to hold with precision better than $10^{-3}$.

\begin{figure}[H]
 \begin{center}
\includegraphics[scale=0.55]{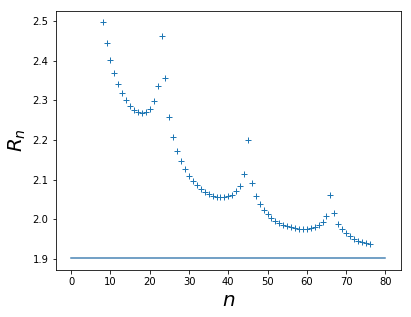} 
\includegraphics[scale=0.55]{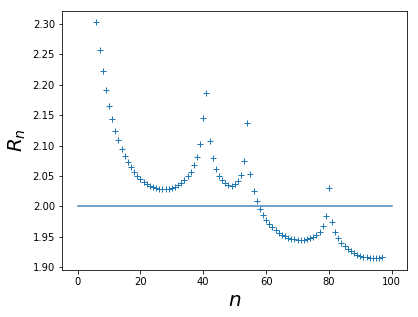} 
\end{center}
\caption{Measured radius of convergence $R_n$, as a function of the order of the truncation of the series $n$, for $\Delta=\cos \pi/5$ (left) and $\Delta=1$ (right). The series converges at $h=0$ if there is only a finite number of points below the  line $\sqrt{h_c}$.}
 \label{fig:mag2}
\end{figure}

However, at larger values of $\Delta$, in particular at $\Delta=1$, the truncated series $\sigma^*(\sqrt{h_c-h})$ keeps oscillating for small magnetic fields with the order of the truncation. Although this is not uncommon in convergent series (the truncated series of $(1+x)^{-1}$ keeps oscillating too near $x=1$), it might also mean that $h_r>0$. To investigate this question further, in Figure \ref{fig:mag2} we plotted $R_n=\exp -\tfrac{\log |a_n|}{n}$ where $a_n$ is the $n$th term of $\sigma^*$ as a series in $\sqrt{h_c-h}$, for $\Delta=\cos\pi/5$ and $\Delta=1$. The series converges at $h=0$ if there is only a finite number of points under the line $\sqrt{h_c}$. This suggests that it is not convergent at $h=0$ for $\Delta$ close to $1$, and that we would have $h_r>0$ at $\Delta=1$.
 
\subsection{An example with complex roots}
Our calculation does not assume that the roots $\lambda_i$ are real. In this section we give the simple example of the Heisenberg spin chain with complex extensive twist $i\varphi$ ($\varphi$ real), where the roots lie on a curve in the complex plane and show that our approach still applies. This variant is for example relevant for the six-vertex model (its free energy $F(m_x,m_y)$ is the Legendre transform of that of this spin chain with respect to $\varphi$ \cite{nohkim}) or for an inhomogeneous version of the Heisenberg spin chain \cite{yungbatchelor,fukuikawakami}:
\begin{equation}
H=\sum_{i=1}^N\frac{1}{2}\left(e^{-\varphi_i}\sigma^+_i\sigma^-_{i+1}+e^{\varphi_i}\sigma^-_i\sigma^+_{i+1} \right)+\sigma_i^z\sigma_{i+1}^z-\tfrac{1}{4}\,,
\end{equation}
where $\sum_{i=1}^N\varphi_i=N\varphi$ are reals. The Bethe equations are
\begin{equation}
s(\lambda_p)=\frac{I_p}{N}+\frac{i\varphi}{\pi}+\frac{1}{N}\sum_{j=1}^kr(\lambda_p-\lambda_j)\,,
\end{equation}
with $s$ and $r$ the same functions as for the untwisted Heisenberg spin chain. As it is, the decomposition of $\lambda_p$ has a non-vanishing $c_{00}$, which implies that there can be some non-vanishing $c_{ab}^{[n]}$ for arbitrary large $n$. To keep the sums over $n$ finite in \eqref{eq:recursive}, we set $\lambda_p'=\lambda_p-s^{-1}(i\varphi/\pi)$ and define
\begin{equation}
\tilde{s}(\lambda)=s(s^{-1}(i\varphi/\pi)+\lambda)-\frac{i\varphi}{\pi}\,,
\end{equation}
so that $c_{00}=0$ and a similar equation to \eqref{eq:recursive} holds. One finds then the free energy with the normalization in \eqref{eq:deffe}
\begin{equation}
\begin{aligned}
F(m)=&-4\cosh^2(\varphi) m +\frac{\pi^2}{3}\cosh(2\varphi) m^3+\frac{\pi^2}{3}(1+\tanh^2\varphi)m^4-\frac{\pi^2 \cosh2\varphi}{60\cosh^4\varphi}\left(\pi^2\cosh^4\varphi -15 \right)m^5\\
&-\frac{\pi^2}{720 \cosh^6\varphi}\left(5\pi^2+(12\pi^2-120)\cosh2\varphi +\pi^2\cosh4\varphi+2\pi^2\cosh6\varphi\right)m^6\\
&+O(m^7)\,.
\end{aligned}
\end{equation}
We stress the fact that the Bethe roots are disposed on a curve in the complex plane that was up to now out of reach of analytic calculations. The agreement with the numerics is still excellent, see Table \ref{tab:xxxtwist} for $\varphi=0.1$ and Table \ref{tab:xxxtwist2} for $\varphi=1$. In this case we do not know the value at $m=1/2$, but we still think the series converges for all $m\leq 1/2$.

\begin{center}
\label{tab:xxxtwist}
\begin{tabular}{|c|c|c|c|c|c|}
\hline
 & $m=0.05$ & $m=0.1$ &$m=0.2$ &$m=0.4$&$m=0.5$\\
 \hline
$N=1000$ & $0.20557908$ & $0.40424331$& $0.77921754$&$1.31659212$&$1.39756987$\\
 \hline
$N=3000$ & $0.20290390$ & $0.40162684$ & $0.77686780$&$1.31553762$&$1.39756837$\\
 \hline
 $N=5000$& $0.20236882$ & $0.40110345$ & $0.77639763$&$1.31532614$&$1.39756824$\\
 \hline
Extrapolation & $0.20156621$ & $0.40031838$ & $0.77569238$&$1.31500891$&$1.39756806$\\
 \hline
$18$-term expansion & $0.20156618$ & $0.40031832$&$0.77569224$ &$1.31500950$&$1.39758483$\\
\hline
\end{tabular}
\captionof{table}{The free energy $-F(m)$ of the twisted XXX spin chain for different magnetizations, for $\varphi=0.1$.}\label{tab:xxxtwist}
\end{center}

\begin{center}
\begin{tabular}{|c|c|c|c|c|c|}
\hline
 & $m=0.05$ & $m=0.1$ &$m=0.2$ &$m=0.4$&$m=0.5$\\
 \hline
$N=1000$ & $0.47464203$ & $0.93959272$& $1.79928056$&$2.95102536$&$3.10980581$\\
 \hline
$N=3000$ & $0.47464147$ & $0.93959159$ & $1.79927833$&$2.95102151$&$3.10980172$\\
 \hline
 $N=5000$& $0.47464143$ & $0.93959150$ & $1.79927815$&$2.95102120$&$3.10980140$\\
 \hline
Extrapolation & $0.47464136$ & $0.93959137$ & $1.79927788$&$2.95102074$&$3.10980091$\\
 \hline
$18$-term expansion & $0.47464140$ & $0.93959145$&$1.79927805$ &$2.95102056$&$3.10978951$\\
\hline
\end{tabular}
\captionof{table}{The free energy $-F(m)$ of the twisted XXX spin chain for different magnetizations, for $\varphi=1$.}\label{tab:xxxtwist2}
\end{center}

\section{Finite-size corrections and critical exponents}
\subsubsection*{Taking into account $N^{-2}$ corrections}
To derive the recurrence relation \eqref{eq:recursive} the thermodynamic limit $N\to\infty$ was taken in \eqref{eq:sum} by keeping only the dominant term. Keeping some sub-dominant terms actually permits to get the finite-size corrections to the free energy. We recall that under the hypothesis of conformal invariance, the $N^{-2}$ corrections to the energy levels contain the Conformal Field Theory (CFT) data such as the central charge $c$ and the conformal exponents $h,\bar{h}$ \cite{bcn,affleck}. Precisely, denoting $e^{\rm gs}_N$ the energy of the ground state in size $N$, and $e_N$ the energy of a given excited state, one has

\begin{equation}
\label{eq:cardy}
e_N^{\rm gs}-e_\infty^{\rm gs}=-\frac{\pi v_F c}{6N^2}+o(N^{-2})\,,\qquad \Delta e_N\equiv e_N-e_N^{\rm gs}=\frac{2\pi v_F}{N^2}(h+\bar{h})+o(N^{-2})\,.
\end{equation}
The finite-size corrections clearly depend on the precise distribution of roots of the state considered, and so the correction terms in \eqref{eq:sum}. For the ground state, i.e. Bethe numbers $-k/2+1/2,...,k/2-1/2$, one has using Faulhaber's formula
\begin{equation}
\label{eq:groundstate}
\frac{1}{N}\sum_{j=1}^k \alpha_j^a=\begin{cases}
\frac{m^{a+1}}{2^{a}(a+1)}-\frac{1}{N^2}\frac{a m^{a-1}}{6\cdot 2^a} +O(N^{-3})\,,\qquad a\, \text{ even}\\
0+O(N^{-3})\,,\qquad a\, \text{ odd}\,.
\end{cases}
\end{equation}
For a state corresponding to a descendant, the Bethe numbers are identical to the ground state expect for $k/2-1/2$ that becomes $k/2+1/2$. Then one has
\begin{equation}
\label{eq:descendant}
\frac{1}{N}\sum_{j=1}^k \alpha_j^a=\begin{cases}
\frac{m^{a+1}}{2^{a}(a+1)}-\frac{1}{N^2}\frac{a m^{a-1}}{6\cdot 2^a}+\frac{1}{N^2}\frac{a m^{a-1}}{2^{a-1}} +O(N^{-3})\,,\qquad a\, \text{ even}\\
\frac{1}{N^2}\frac{a m^{a-1}}{2^{a-1}} +O(N^{-3})\,,\qquad a\, \text{ odd}\,.
\end{cases}
\end{equation}
For a so-called "Umklapp" state, $-k/2+1/2$ is replaced by $k/2+1/2$. Thus
\begin{equation}
\frac{1}{N}\sum_{j=1}^k \alpha_j^a=\begin{cases}\frac{m^{a+1}}{2^{a}(a+1)}-\frac{1}{N^2}\frac{a m^{a-1}}{6\cdot 2^a}+\frac{1}{N^2}\frac{a m^{a-1}}{2^{a-1}} +O(N^{-3})\,,\qquad a\, \text{ even}\\
\frac{1}{N}\frac{m^{a}}{2^{a-1}} +O(N^{-3})\,,\qquad a\, \text{ odd}\,.
\end{cases}
\end{equation}

With this, a recurrence relation analogous to \eqref{eq:recursive} with $N^{-1}$ and $N^{-2}$ corrections can be derived. However, the sum $a_1+b_1+b_2+1=b$ is replaced by $a_1+b_1+b_2-1=b$ for $N^{-2}$ corrections for example, and $c_{ab}$ could appear in the right-hand side of the recurrence relation in the term proportional to $N^{-2}$. Even if it does, it does not prevent from computing recursively the expansion of $c_{ab}$ to a desired order: one can just compute the $c_{ab}$'s order by order in $N^{-1}$. In fact, closer inspection indicates that it does not happen for the states discussed above.
%

The recurrence relation \eqref{eq:recursive} with a $N^{-2}$ term becomes e.g. for the ground state
\begin{equation}
\begin{aligned}
\label{eq:recursiveGS}
c_{ab}=&\frac{1}{s'(0)}\left(\sum_{\substack{a_1+b_1+b_2+1=b\\ a_1 \text{ even}}} \left(\sum_{n\geq 0}\frac{r^{(n)}(0)}{n!}\sum_{q=0}^{n}(-1)^q {n \choose q}\frac{c_{a_1b_1}^{[q]} c_{ab_2}^{[n-q]}}{2^{a_1} (a_1+1)}\right)-\sum_{n\geq 1} \frac{c^{[n]}_{ab}s^{(n)}(0)}{n!} \right.\\
&-\left. \frac{1}{6N^2}\sum_{\substack{a_1+b_1+b_2-1=b\\ a_1 \text{ even}}} \left(\sum_{n\geq 0}\frac{r^{(n)}(0)}{n!}\sum_{q=0}^{n}(-1)^q {n \choose q}a_1\frac{c_{a_1b_1}^{[q]} c_{ab_2}^{[n-q]}}{2^{a_1}}\right) \right)\,.
\end{aligned}
\end{equation}

See Table \ref{tab:xxxn} for a numerical verification of these relations in size $N=200$ for the XXX spin chain, for the ground state (GS), a descendant state (D) and the Umklapp state (U).
\begin{center}
\begin{tabular}{|c|c|c|c|c|}
\hline
 & $m=0.05$ & $m=0.1$ &$m=0.2$ &$m=0.4$\\
 \hline
GS(num.)& $0.1995722665$ & $0.3963828609$& $0.7682525850$&$1.303613450$\\
 \hline
GS (th.) & $0.1995722664$ & $0.3963828606$ & $0.7682525844$&$1.303614161$\\
 \hline
 \hline
D (num.) & $0.1995206015$ & $0.3962754120$& $0.7680261945$&$1.303174308$\\
 \hline
D (th.) & $0.1995205738$ & $0.3962753814$ & $0.7680261588$&$1.303170391$\\
 \hline
 \hline
U (num.) & $0.1995231282$ & $0.3962858646$ & $0.7680691881$&$1.303332830$\\
 \hline
U (th.) & $0.1995231262$ & $0.3962858613$ & $0.7680691846$&$1.303329318$\\
\hline
\end{tabular}
\captionof{table}{Some energy levels of the Heisenberg spin chain for different magnetizations, in finite size $N=200$.}\label{tab:xxxn}
\end{center}
\subsubsection*{Spectrum of the model}
Let us denote $\Delta_{\rm GS}, \Delta_{\rm D}, \Delta_{\rm U}$ the $N^{-2}$ correction to the three states previously described. We define
\begin{equation}
\label{eq:gg}
\begin{aligned}
v_F&=\frac{\Delta_{\rm D}-\Delta_{\rm GS}}{2\pi}\,,\qquad c=-\frac{6\Delta_{\rm GS}}{\pi v_F}\,,\\
g&=\frac{F''(m)}{\pi v_F}\,,\qquad g^*=\frac{\Delta_{\rm U}-\Delta_{\rm GS}}{2\pi v_F}\,,
\end{aligned}
\end{equation}
and let us consider a state obtained from the ground state by removing $n$ roots, changing the sign of $n^*/2$ negative roots (that become thus positive -- a negative $n^*$ means that the sign of $|n^*|/2$ positive roots is changed), and globally shifting the Bethe integers of the roots by $p_\pm$ for positive/negative roots (i.e., the sum of the positive Bethe integers is increased by $p_+$ and the sum of the negative Bethe integers by $-p_-$). We assume that the Bethe integer configuration is still a valid configuration, and that the modifications are done only at finite distance from the maximal and minimal roots of the ground state.\\

The $n$ direction can actually be simply taken into account by changing $m$ into $m-n/N$, so that one can assume in a first step $n=0$. In this case one gets, with $\sum_{i=0}^{n^*-1}(1/2+i)=(n^*)^2/2$
\begin{equation}
\frac{1}{N}\sum_{j=1}^k \alpha_j^a=\begin{cases}\frac{m^{a+1}}{2^{a}(a+1)}-\frac{1}{N^2}\frac{a m^{a-1}}{6\cdot 2^a}+\frac{(n^*)^2}{N^2}\frac{a m^{a-1}}{2^{a+1}} +\frac{p_++p_-}{N^2}\frac{a m^{a-1}}{2^{a-1}} +O(N^{-3})\,,\qquad a\, \text{ even}\\
\frac{1}{N}\frac{m^{a}}{2^{a}}n^*+\frac{p_+-p_-}{N^2}\frac{a m^{a-1}}{2^{a-1}}  +O(N^{-3})\,,\qquad a\, \text{ odd}\,.
\end{cases}
\end{equation}
Remark that the term with $a$ odd can appear in the result only through its square, since it is multiplied by $-1$ if all the roots are multiplied by $-1$. It follows that there is no $N^{-1}$ correction, that the $n^*/N$ correction appears in the result through its square, and that the $p_\pm$ term for $a$ odd does not contribute to the result. Thus the $N^{-2}$ correction is $\Delta_{\rm GS}+(\Delta_{\rm D}-\Delta_{\rm GS})(p_++p_-)+(\Delta_{\rm U}-\Delta_{\rm GS})(n^*/2)^2$.\\
As for the $n$ direction, it gives the correction to the free energy $F(m)-\frac{n}{N}F'(m)+\frac{n^2}{2N^2}F''(m)$. The $N^{-1}$ term is compensated by the magnetic field part in the energy of the chain, so that remains only the $N^{-2}$ part. It follows that the $N^{-2}$ correction $\Delta_{\rm Gen}$ of this generic state is
\begin{equation}
\Delta_{\rm Gen}=2\pi v_F\left(-\frac{c}{12}+\frac{n^2}{4} g+ \frac{(n^*)^2}{4} g^*+p_++p_- \right)\,.
\end{equation}
Note that $c=1$ from \eqref{eq:descendant} and \eqref{eq:groundstate}. To fully recover the spectrum of a free compact boson, one needs
\begin{equation}
\label{eq:ging}
g^*=\frac{1}{g}\,,
\end{equation}
which is a very non-trivial statement on the recurrence relations for $c_{ab}$, that we could not prove at all orders, but checked order by order in $m$ until $m^{28}$. The fact that $g$ and $1/g^*$ give the same expansion is a very non-trivial check of validity of our approach.

For example one has for the XXZ spin chain
\begin{equation}
\begin{aligned}
g=&1+\frac{2\Delta}{1+\Delta}m+\frac{3\Delta^2}{(1+\Delta)^2}m^2+\frac{\Delta}{3(1+\Delta)^3}(-\pi^2+\pi^2\Delta+12\Delta^2)m^3\\
&+\frac{5\Delta^2}{6(1+\Delta)^4}(-2\pi^2+2\pi^2\Delta+6\Delta^2)m^4\\
&+\frac{1}{60(1+\Delta)^5}\left(360\Delta^5-300\Delta^3\pi^2+300\Delta^4\pi^2+\Delta\pi^4-11\Delta^2\pi^4+11\Delta^3\pi^4-\Delta^4\pi^4 \right)m^5\\
&+O(m^6)\,,
\end{aligned}
\end{equation}
and the Luttinger parameter $K=1/g$:
\begin{equation}
\begin{aligned}
K&=1-\frac{2\Delta}{1+\Delta} m +\frac{\Delta^2}{(1+\Delta)^2}m^2-\frac{(\Delta-1)\Delta \pi^2}{3(1+\Delta)^3}m^3-\frac{(\Delta-1)\Delta^2\pi^2}{3(1+\Delta)^4}m^4\\
&+\frac{(\Delta-1)\Delta \pi^2(\pi^2-10\Delta \pi^2-20\Delta^2 +\Delta^2\pi^2)}{60(1+\Delta)^5}m^5\\
&+\frac{1}{60(1+\Delta)^6}\left(20\Delta^4\pi^2-20\Delta^5\pi^2-2\Delta^2\pi^4+35\Delta^3\pi^4-38\Delta^4\pi^4+3\Delta^5\pi^4 \right) m^6\\
&+O(m^7)\,.
\end{aligned}
\end{equation}

\subsubsection*{Critical exponent of the spin-spin two-point function}
\begin{figure}[H]
 \begin{center}
\includegraphics[scale=0.56]{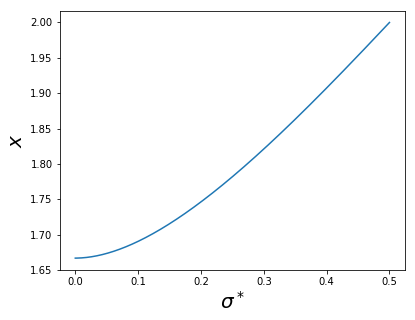} 
\includegraphics[scale=0.56]{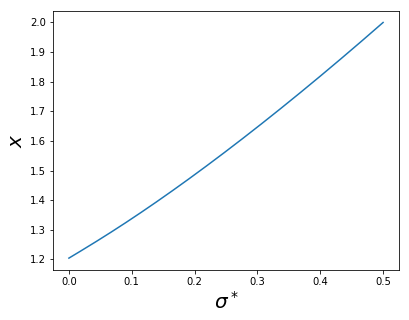} 
\end{center}
\caption{Critical exponent $x$ as a function of the magnetization $\sigma$, for $\Delta=\cos2\pi/5$ (left) and $\Delta=\cos\pi/6$ (right), using a $28$-term expansion in $m$.}
 \label{fig:crit}
\end{figure}

The spin-spin correlations decay with distance $r$ as $r^{-x}$ with $x=2/g$ for $0\leq \Delta \leq 1$ \cite{korepin}. This yields the critical exponent
\begin{equation}
\begin{aligned}
x&=2-\frac{4\Delta}{1+\Delta} m +\frac{2\Delta^2}{(1+\Delta)^2}m^2-\frac{2(\Delta-1)\Delta \pi^2}{3(1+\Delta)^3}m^3-\frac{2(\Delta-1)\Delta^2\pi^2}{3(1+\Delta)^4}m^4\\
&+\frac{(\Delta-1)\Delta \pi^2(\pi^2-10\Delta \pi^2-20\Delta^2 +\Delta^2\pi^2)}{30(1+\Delta)^5}m^5\\
&+\frac{1}{30(1+\Delta)^6}\left(20\Delta^4\pi^2-20\Delta^5\pi^2-2\Delta^2\pi^4+35\Delta^3\pi^4-38\Delta^4\pi^4+3\Delta^5\pi^4 \right) m^6\\
&+O(m^7)\,.
\end{aligned}
\end{equation}
Again, only the first term of this expansion was known \cite{bogoliubovizerginkorepin,korepin}. In the case of a spin-$1$ chain in a magnetic field, the first two terms of a similar series were computed by different means \cite{konikfendley}. See Figure \ref{fig:crit} for a plot of this expansion for different values of $\Delta$. At $\sigma=0$ we know $x=(1-\frac{1}{\pi}\arccos\Delta)^{-1}$ \cite{devegawoynarovich,woynarovicheckle,karowski,klumperbatchelor,klumper}. With a $28$-term expansion, we find for $\Delta=\cos2\pi/5$, $x=1.66663$ (exact value $5/3$), and $x=1.204$ for $\Delta=\cos\pi/6$ (exact value $6/5$). For $m\to 0$, i.e. close to the critical magnetic field $h_c$, the critical exponent goes to that of the free fermion case $\Delta=0$, independently of $\Delta$.

\section{Conclusion}
In this paper we computed the free energy of the Heisenberg and XXZ spin chains, as a series in $1/2-\langle \sigma^z_i\rangle$ where $-1/2\leq \langle \sigma^z_i\rangle \leq 1/2$ is the mean magnetization per site. All the terms of the series can be computed in closed form recursively with only algebraic manipulations. The series is numerically shown to be convergent in the full physical range $0\leq m\leq 1/2$. Such an expansion is new. The previous attempts to solve this problem seem to have essentially resorted to an intractable Wiener-Hopf equation.

By taking the Legendre transform of this series, we obtain the acquired magnetization of the chain $\sigma$ in a magnetic field $h$ as a series in $\sqrt{h_c-h}$ where $h_c$ is the smallest field for which it is fully polarized. This expansion converges in a range $h_r<h\leq h_c$ where $h_r$ depends on $\Delta$, is equal to $-h_c$ for $\Delta=0$ and is negative at least for $\Delta \lessapprox 0.6$. For $\Delta$ close to $1$, the numerics suggests that $h_r>0$ (i.e., the series does not converge at $h=0$), although one needs more than $50$ terms to observe it. At least within the radius of convergence of the series, this solves the problem first addressed in \cite{griffiths}, that is the response of the Heisenberg and XXZ quantum spin chains to an external magnetic field.

To derive these results, we expressed the free energy of the spin chain at mean magnetization per site $\langle \sigma^z_i\rangle$ as a series in $1/2-\langle \sigma^z_i\rangle$ whose coefficients can be computed in closed form through a recurrence equation. This is done by reformulating the Bethe equations as a recurrence relation on the coefficients of their roots when written as a double series in their Bethe numbers and $1/2-\langle \sigma^z_i\rangle$.

The same reasoning permits to derive a similar series for the finite-size corrections of the free energy, that are related to the spectrum of the field theory that describes the spin chain in the continuum limit, upon the hypothesis of conformal invariance. We checked that it has the form of that of a free compactified boson, whose radius can be expressed as a series as well. From this we deduced the critical exponent of the spin-spin correlation functions as a series in $1/2-\langle \sigma^z_i\rangle$.

This novel approach seems very promising, both for its simplicity and its power. For example we showed that it also works for a case with roots lying on a curve in the complex plane, previously out of reach of analytical calculations. Natural extensions of this method would be to generalize it to nested Bethe ansatz models or quantum transfer matrices. It opens many possibilities, such as the study of phase transitions in quantum spin chains.

\smallskip
\noindent{\bf Acknowledgments}: this work was supported in part by the Advanced ERC Grant NuQFT. E.G. thanks F. Essler and P. Fendley for interesting discussions.

\smallskip

\appendix

\subsection*{Appendix: A sum over trees}
It is possible to express the solution to the recurrence relation \eqref{eq:recursive} as a sum over trees. We can first rewrite \eqref{eq:recursive} in terms of partitions of $a$ and $b$:
\begin{equation}
\begin{aligned}
c_{ab}&=\frac{1}{s'(0)}\left(\sum_{\substack{a_1+...+a_{n-p}=a\\ b_1+...+b_{n+p}=b-1}}(-1)^p\frac{r^{(n)}(0)}{p!(n-p)!}\frac{\pi(x_{\{b\}})}{2^{x_{\{b\}}}(x_{\{b\}}+1)}\prod_{i=1}^{n-p}c_{a_ib_i}\prod_{i=1}^{p}c_{b_{n-p+2i-1}b_{n-p+2i}} \right.\\
&\left.-\sum_{\substack{a_1+...+a_{n}=a\\ b_1+...+b_n=b\\n\geq 2}}\frac{s^{(n)}(0)}{n!}\prod_{i=1}^nc_{a_ib_i} \right)\,.
\end{aligned}
\end{equation}

The number $x_{\{b\}}$ is $b_{n-p+1}+b_{n-p+3}+...+b_{n+p-1}$. $\pi(x)$ is $0$ if $x$ is odd, and $1$ if it is even. Applying recursively this equation naturally leads to a formulation of the result in terms of trees. The $s$ terms will be represented by a node of type $s$ with $n$ descendants, and the $r$ terms by a node of type $r$ with $n$ descendants among which $p$ are marked. The constraints for the $s$ and $r$ nodes $a_1+...+a_n=a$ and $a_{1}+...+a_{n-p}=a$ mean that there must be $a$ leaves connected to the root by a path without marked edges. The constraints $b_1+...+b_{n}=b$ for a $s$ node and $b_1+...+b_{n+p}=b-1$ for a $r$ node mean that there are $a+b-\#r$ leaves in total, where $\#r$ is the number of $r$ nodes in the tree. The number $b_{n-p+1}+b_{n-p+3}+...+b_{n+p-1}$ at a node of type $r$ is given by the number of leaves that are connected to one of its descendants with marked edge by a path without marked edges. To each node is associated a weight $w$ that is
\begin{equation}
\begin{aligned}
w_s&=-\frac{1}{s'(0)}\frac{s^{(n)}(0)}{n!}\,,\qquad \text{for a node of type $s$ with $n\geq 2$ descendants (zero for $n=1$)}\\
w_r&=(-1)^p\frac{1}{s'(0)}\frac{r^{(n)}(0)}{n!}\frac{\pi(x)}{2^{x}(x+1)}\,,\\
&\qquad \text{for a node of type $r$ with $n$ descendants among which $p$ are marked}\,,
\end{aligned}
\end{equation}
where $x$ is the number of leaves connected to one of the descendants with marked edge of the node by a path without marked edges. The weight of a tree is given by the product of the weights of its nodes, times $1/s'(0)$ for each of its leaves.

The coefficient $c_{ab}$ is given by the sum of the weights of all such trees with $\#r$ nodes $r$ and with $a+b-\#r$ leaves and $a$ leaves connected to the root by a path without marked edges, where $\#r$ has to be summed over.

Denote $\mathcal{T}_a$ the set of trees with at most $a-1$ leaves and $a-1$ $r$ nodes, such that the root is a $s$ node with an even number of descendants (the other nodes can be similarly $s$ nodes with $\geq 2$ descendants, or $r$ nodes, and the edges descendant of $r$ nodes can be marked). The weight $w$ of a tree in $\mathcal{T}_a$ is given by $w_s$ for a $s$ node that is not the root, $w_r$ for a $r$ node, and $w_{\rm root}$ for the root with
\begin{equation}
w_{\rm root}=\frac{s^{(n+1)}(0)}{n!}\frac{\pi(x)}{2^{x}(x+1)} \qquad \text{if the root has $n$ descendants}\,,
\end{equation}
where $x$ is the number of leaves connected to the root by a path without marked edges. The free energy coefficient $f_a$ is $-2\pi$ times the sum of the weights of the trees in $\mathcal{T}_a$.\\

We also mention that the generating function of the $c_{ab}$'s (in the $a$ direction) defined by these trees is shown without problems to satisfy an equation that is the Bethe equations written with the so-called counting function $z$ (the integral of the density) in the thermodynamic limit. This could potentially enter an alternative proof of the condensation of Bethe roots in the thermodynamic limit \cite{kozlowski}.
\bibliography{expansion}
\bibliographystyle{ieeetr}

\end{document}